\documentclass{aa}  

\usepackage{graphicx}
\usepackage[caption=false]{subfig}
\usepackage[colorlinks=true, allcolors=blue]{hyperref} 

\usepackage{txfonts}
\begin{document} 
    
    \title{Outflows, cores, and magnetic field orientations in W43-MM1 as seen by ALMA}
   \author{C. Arce-Tord\inst{1}
          \and
           F. Louvet\inst{1,2}
          \and
           P. C. Cortes\inst{3, 4}
           \and
           F. Motte\inst{5}
         \and
         C. L. H. Hull\inst{3,6,7}
         \and
          V. J. M. Le Gouellec\inst{8,2}
         \and
          G. Garay\inst{1}
          \and
          T. Nony\inst{5}
          \and
         P.\,\,Didelon\inst{2}    
         \and
         L. Bronfman\inst{1}
          }
   \institute{Universidad de Chile, Camino el Observatorio 1515, Las Condes, Santiago de Chile
              \email{carce@das.uchile.cl}
        \and
Laboratoire AIM Paris-Saclay, CEA/IRFU – CNRS/INSU – Universit\'{e} Paris Diderot, Service d'Astrophysique, B\^{a}t. 709,CEA-Saclay, 91191 Gif-sur-Yvette Cedex, France         
\email{fabien.louvet@cea.fr}
\and
            Joint ALMA Observatory, Alonso de Cordova 3107, Vitacura, Santiago, Chile
        \and
             National Radio Astronomy Observatory, Charlottesville, VA 22903, USA
        \and
             Univ. Grenoble Alpes, CNRS, IPAG, F-38000 Grenoble, France
        \and
           National Astronomical Observatory of Japan, NAOJ Chile, Alonso de C\'ordova 3788, Office 61B, 7630422, Vitacura, Santiago, Chile
        \and
            NAOJ Fellow,
        \and
        European Southern Observatory, Alonso de Córdova 3107, Vitacura, Casilla 19001, Santiago , Chile 
             }


 
  \abstract
   {}
  {It has been proposed that the magnetic field, which is pervasive in the interstellar medium (ISM), plays an important role in the process of massive star formation. To better understand the impact of the magnetic field at the pre- and protostellar stages, high-angular resolution observations of polarized dust emission toward a large sample of massive dense cores are needed. We aim to reveal any correlation between the magnetic field orientation and the orientation of the cores and outflows in a sample of protostellar dense cores in the W43-MM1 high-mass star-forming region.}
   {We used the Atacama Large Millimeter Array in Band 6 (1.3\,mm) in full polarization mode to map the polarized emission from dust grains at a physical scale of $\sim$2700\,au. We used these data to measure the orientation of the magnetic field at the core scale. Then, we examined the relative orientations of the core-scale magnetic field, of the protostellar outflows, and of the major axis of the dense cores determined from a 2D Gaussian fit in the continuum emission.}
   {We find that the orientation of the dense cores is not random with respect to the magnetic field. Instead, the dense cores are compatible with being oriented 20-50$^\circ$ with respect to the magnetic field. As for the outflows, they could be oriented 50-70$^\circ$ with respect to the magnetic field, or randomly oriented with respect to the magnetic field, which is similar to current results in low-mass star-forming regions.}
   {The observed alignment of the position angle of the cores with respect to the magnetic field lines shows that the magnetic field is well coupled with the dense material; however, the 20-50$^\circ$ preferential orientation contradicts the predictions of the magnetically-controlled core-collapse models. The potential correlation of the outflow directions with respect to the magnetic field suggests that, in some cases, the magnetic field is strong enough to control the angular momentum distribution from the core scale down to the inner part of the circumstellar disks where outflows are triggered.}

\keywords{ISM: magnetic fields -- ISM: clouds -- ISM: jets and outflows -- Stars: formation -- Submillimeter: ISM -- Techniques: interferometric}
   \maketitle
%

\section{Introduction}

It has been proposed that the magnetic field, which is pervasive in the interstellar medium (ISM), might play an important role in the dynamical evolution of star-forming clouds \citep[e.g.,][]{shu+87, hennebelle, commercon11, crutcher+12, planck16_bfield_molecular_clouds, planck16_ismfilaments,  matsushita18, beuther18}. Large-scale observations ($\sim$1\,pc) of the magnetic field revealed a well-ordered structure in the low-density envelopes of molecular clouds, suggesting that parsec-scale envelopes are magnetically supported against gravitational collapse \citep[e.g.,][]{franco10}. Besides their possible role in supporting clouds against gravity, magnetic fields may also strongly affect the formation and evolution of substructures within the clouds. Indeed the magnetic fields sometimes have a consistent morphology throughout scales, as reported by the TADPOL survey \citep{hull+14}, which compares the orientation of the magnetic field at 20$^{''}$ and 2.5$^{''}$ in 30 low-mass star-forming regions distant of 125 to 2650\,pc. At core scales ($\sim$\,0.1\,pc), the magnetically-regulated core-collapse models presume a dominant role of the magnetic field \citep{shu+87, shu+04, galli93a,tomisaka98, allen+03b, allen+03a}. These models are consistent with a handful of observations \citep[e.g.,][]{chapman13, qiu+14}, but due to a small fraction of consistent data, the magnetically dominant core-collapse does not seem to be the predominant mode of low- or high-mass star formation \citep{hullrev_19}. At the circumstellar disk scales, $\sim$\,200\,au, it is predicted that the magnetic fields have a perpendicular component to the disks, thus having a leading role for the triggering of outflows \citep{bland-payne, camenzind90,konigl2000, shu2000}. No systematic relation, however, has been observed thus far between the magnetic field direction and the outflow orientation when these are probed in the low-mass regime at $\sim$1000\,au scales \citep[][]{hull+13, hullrev_19}. Nevertheless, one low-mass study toward 12 Class 0 protostars by \cite{galametz18} shows that at the envelope scale (600-1500\,au), the magnetic field is preferentially oriented either parallel or perpendicular to the direction of the outflows. 

A few interferometric studies, such as \citet[][]{girart+09}, \citet[][]{beuther10}, and \citet[][]{sridha+14}, have probed the high-mass star-forming regime. In particular, \citet{zhang14} performed multi-scale observations of the magnetic field orientation in 14 high-mass star-forming regions. They show that the magnetic field at core scales is either perpendicular or parallel to the magnetic field orientation at clump scales. Similar results have also been reported by \citet{koch+14} toward 50 star-forming regions, and by \citet{ching+17} toward six massive dense cores located in DR21. One issue in determining the importance of magnetic fields for high-mass star formation is the lack of consistent analysis, in a large sample of sources, of the magnetic field morphology with respect to the density structure and gas dynamics. We collected all of this information in the high-mass star-forming region (HMSFR) W43-MM1. W43-MM1, which is at a distance of 5.5\,kpc from the Sun \citep{zhangb+14}, has been reported as a mini-starburst cluster due to its high star formation rate of $\sim$6000 M$_{\odot}/{\rm Myr}$ \citep{louvet+14}. W43-MM1 has been observed in polarized dust emission at 1.3\,mm by \citet{cortes-crutcher} with BIMA. They reported an ordered polarization pattern at $\sim$4.5$^{''}$ angular resolution and derived an on-the-plane of the sky magnetic field strength of 1.7\,mG. Later, \citet{sridha+14} observed W43-MM1 in polarized dust emission at 345\,GHz and an angular resolution of $\sim$2.3$^{''}$ with the SMA, and they report a pinched morphology of the magnetic field with a field strength of $\sim$6\,mG. More recently, \citet{cortes+16} conducted the first ALMA high-angular resolution map at 0.5$^{''}$ of the magnetic fields in W43-MM1, which was performed using a single pointing, and they also report an organized magnetic field morphology with field strengths ranging from 0.2 to 9\,mG.

In this article, we present observations of the polarized dust emission toward W43-MM1 in five pointings with the ALMA interferometer. We compare the orientation of the plane-of-the-sky magnetic field with the orientation of the cores and outflows identified in W43-MM1. \cite{mottenat18} report that about 130 dense cores were forming stars in this region. From this sample of cores, 27 are driving 46 outflow lobes, whose orientations have been traced by \cite{nony20} using CO(2-1) and SiO(5-4) emission lines with ALMA. In Sect.\,\ref{s:obs} we describe the observations, calibration, and imaging of our data. In Sect.\,\ref{s:results}, we calculate the angle differences among the magnetic field, the orientations of the outflows, and the position angle (PA) of the cores; additionally, we build their statistical distributions in the form of cumulative distribution functions (CDFs) that are further compared with synthetic populations of angle distributions. In Sect.\,\ref{s:diss} we discuss the physical implications of our results and conclude in  Sect.\,\ref{conc}.


\section{Observations}\label{s:obs}

\begin{figure*}[!ht]
    \centering
     \subfloat{\includegraphics[trim =0cm 0cm 0cm 0cm, width=1.0\linewidth]{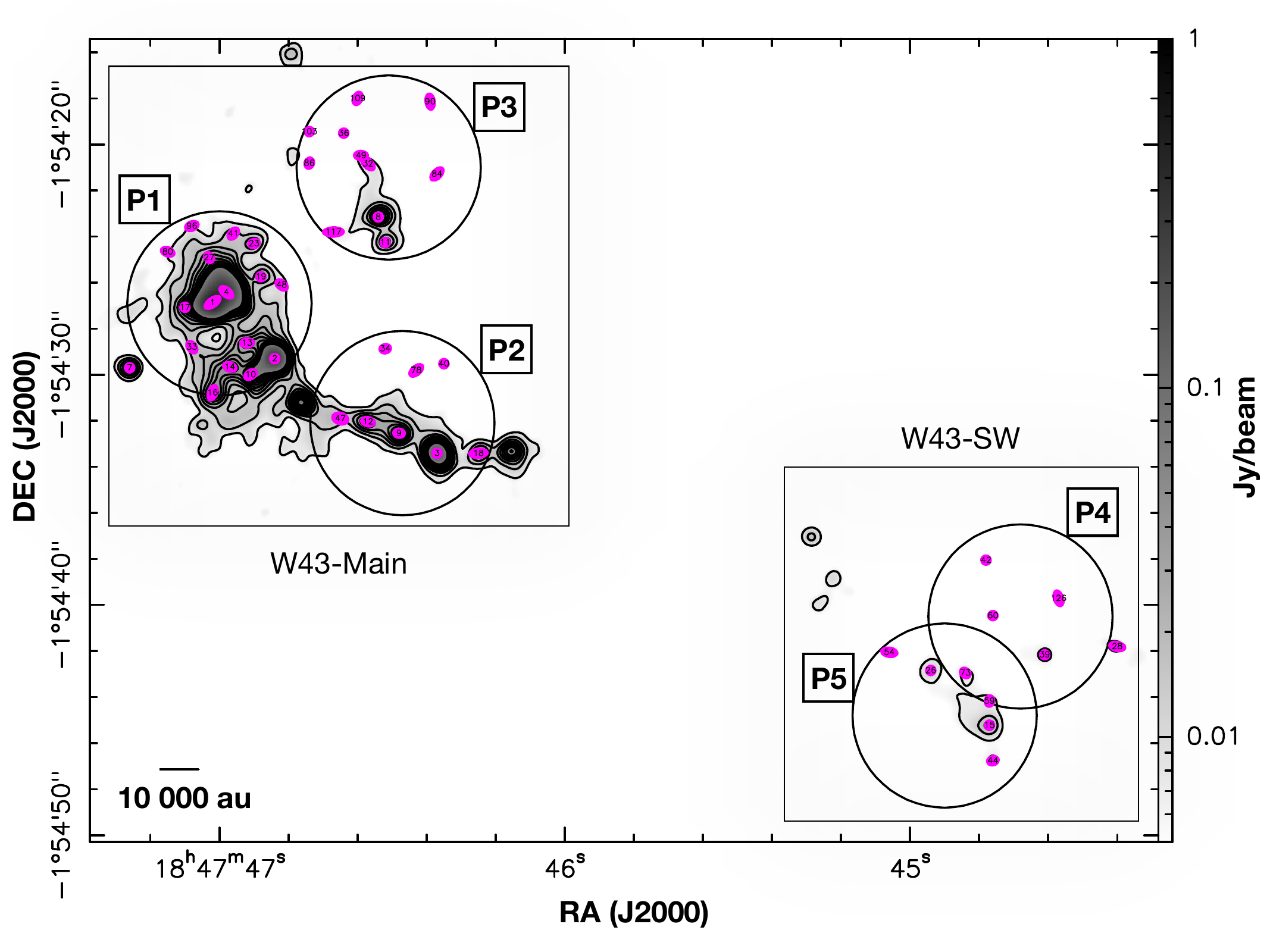}}
    \caption{Continuum emission at 1.3\,mm of the W43-MM1 HMSFR. Contours start at 3$\sigma$ with steps of 3$\sigma$, where $\sigma$=2.4\,mJy/beam. Black circles illustrate one third of the primary beam for each of the five pointings, which are further presented in Figs.~\ref{f:pointings_1} and \ref{f:pointings_2}. The magenta ellipses indicate the cores presented by \cite{mottenat18}. }
     \label{f:overview}
\end{figure*}

\begin{figure*}[!ht]
    \centering
     \subfloat[Pointing 1]{\label{fig:1}\includegraphics[trim=2cm 0cm 28.5cm 0cm, width=0.35\linewidth]{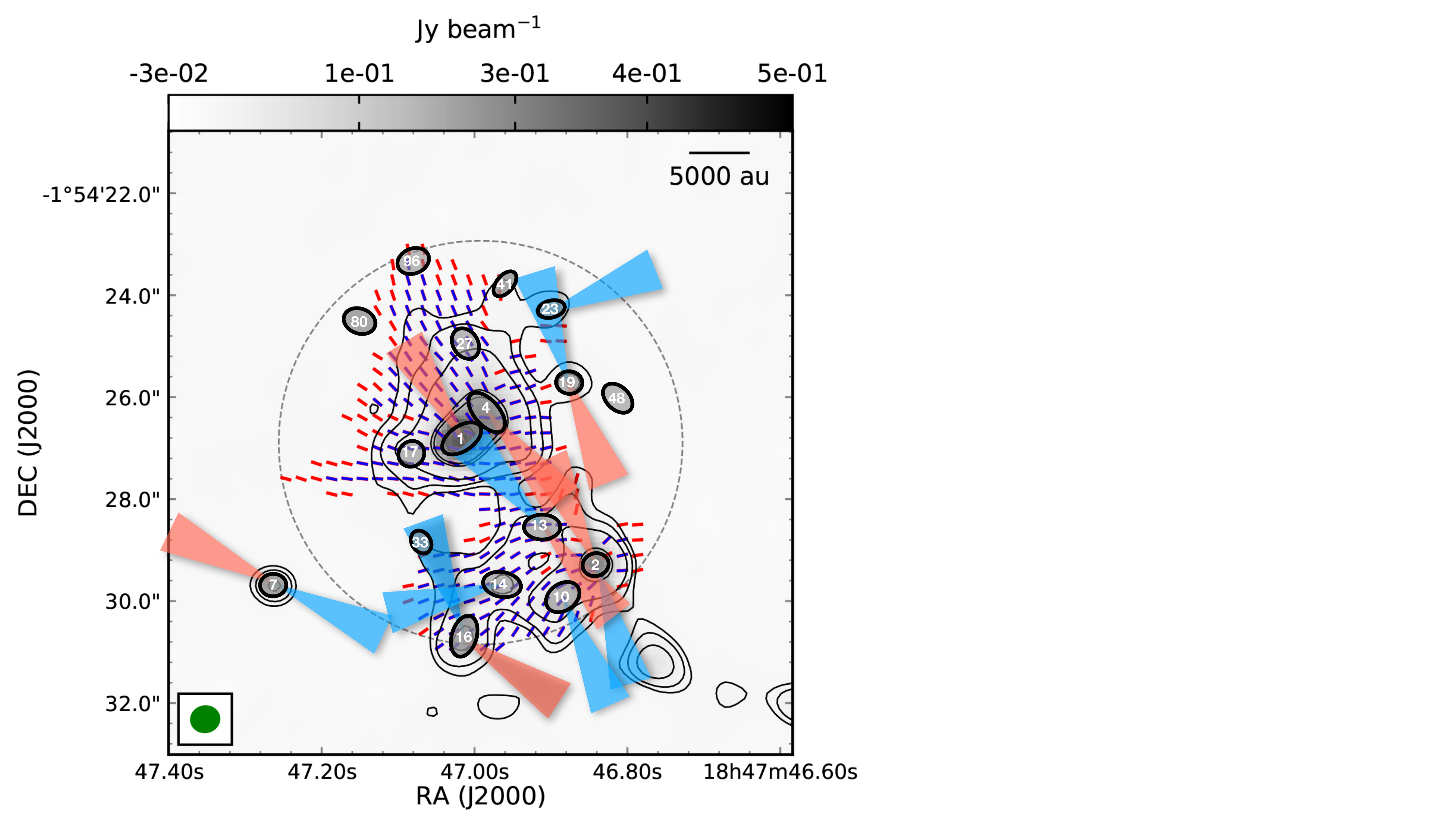}}
    \subfloat[Pointing 2]{\label{fig:2}\includegraphics[trim=2cm 0cm 28.5cm 0cm, width=0.35\linewidth]{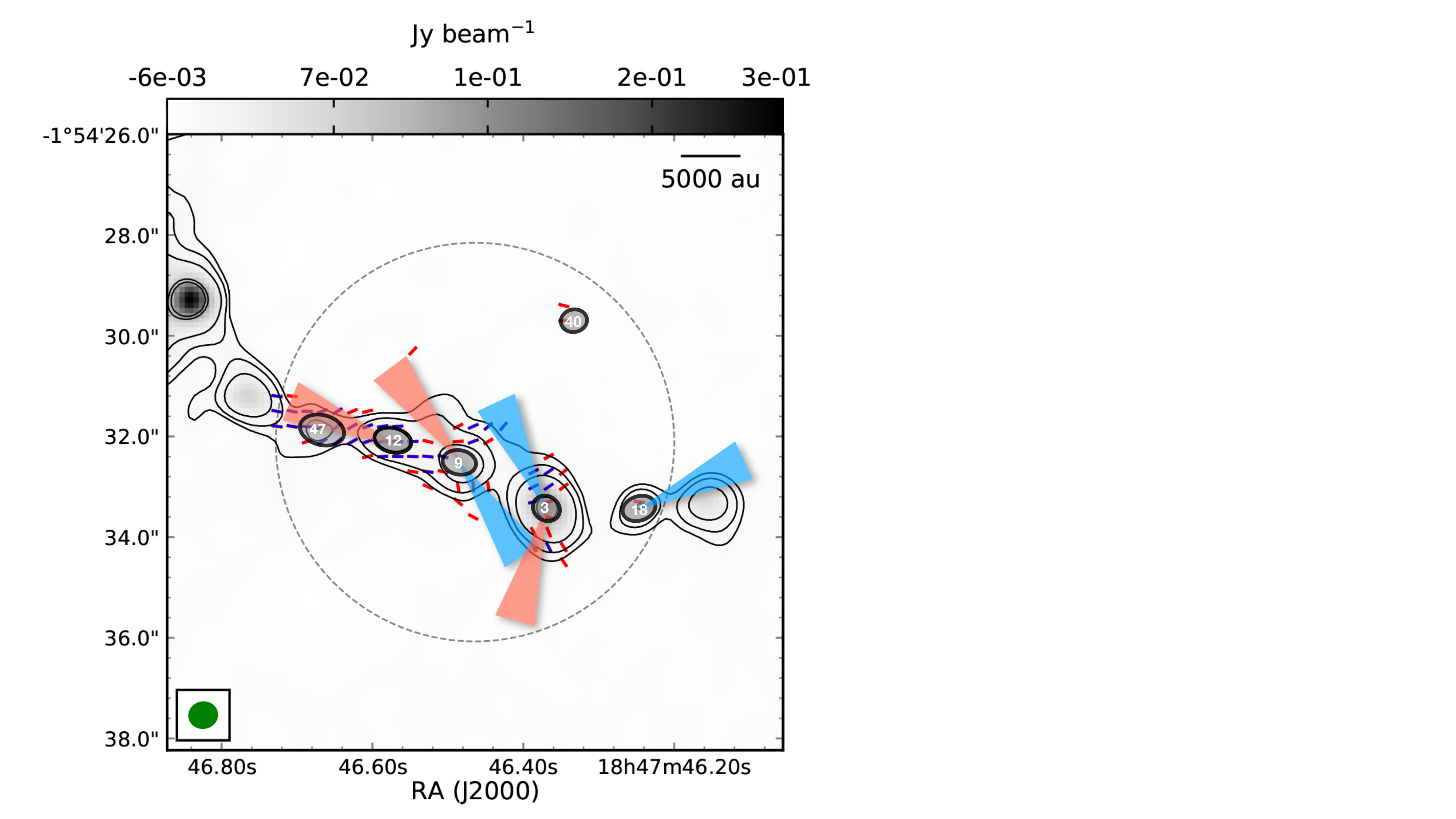}}
    \subfloat[Pointing 3]{\label{fig:3}\includegraphics[trim=2cm 0cm 28.5cm 0cm, width=0.35\linewidth]{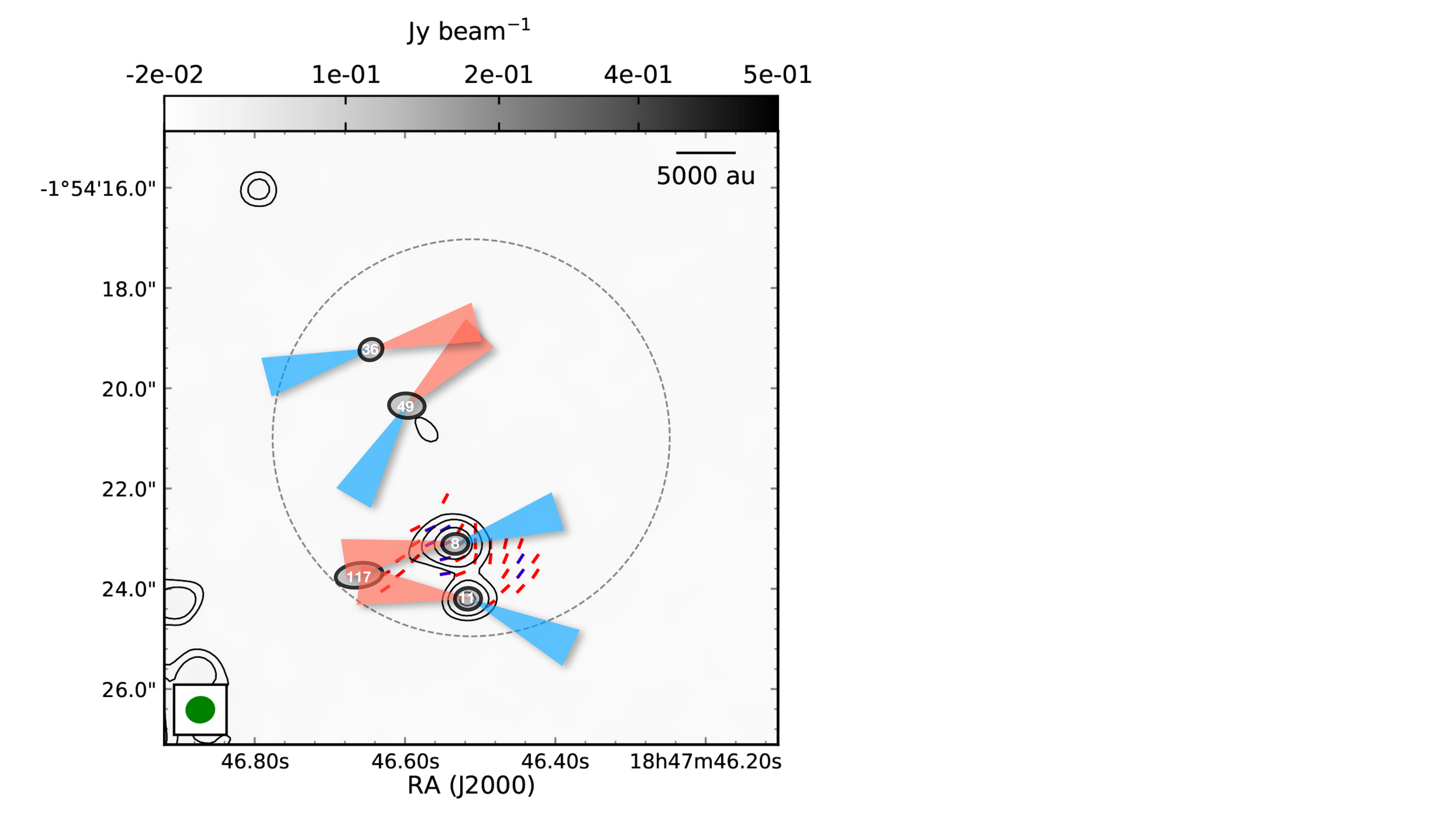}}
    \caption{Dust polarization semi-vectors (rotated by 90$^\circ$ to show the inferred magnetic field orientation) over the continuum emission at 1.3\,mm for pointings 1, 2, and 3 toward W43-MM1. Continuum contours levels are at 3, 5, 10, 50, and 100~$\sigma_{\rm I}$ ($\sigma_{\rm I}$ levels in Table\,\ref{t:rms}). The semi vectors in red and blue show the magnetic field orientation where the polarized intensity exceeds a noise level of 3$\sigma_{\rm P}$ and 5$\sigma_{\rm P}$, respectively (see Table\,\ref{t:rms}). The semi-vectors are plotted every three pixels, which correspond to a Nyquist spatial frequency of four vectors per synthesized beam (two in each dimension). The red and blue cones represent the red-shifted and blue-shifted outflow lobes, respectively. Dashes gray circles represent one third of the $\sim$\,24$^{\prime\prime}$ primary beam, within which we performed the analysis. The solid green ellipse shows the synthesized beam of $0.55^{\prime\prime}\times 0.49^{\prime\prime}$, PA\,=\,-79.4$^{\circ}$. }
    \label{f:pointings_1}
\end{figure*}

\begin{figure*}[!ht]
    \centering
    \subfloat[Pointing 4]{\label{fig:4}\includegraphics[trim=2cm 0cm 28.5cm 0cm, width=0.35\linewidth]{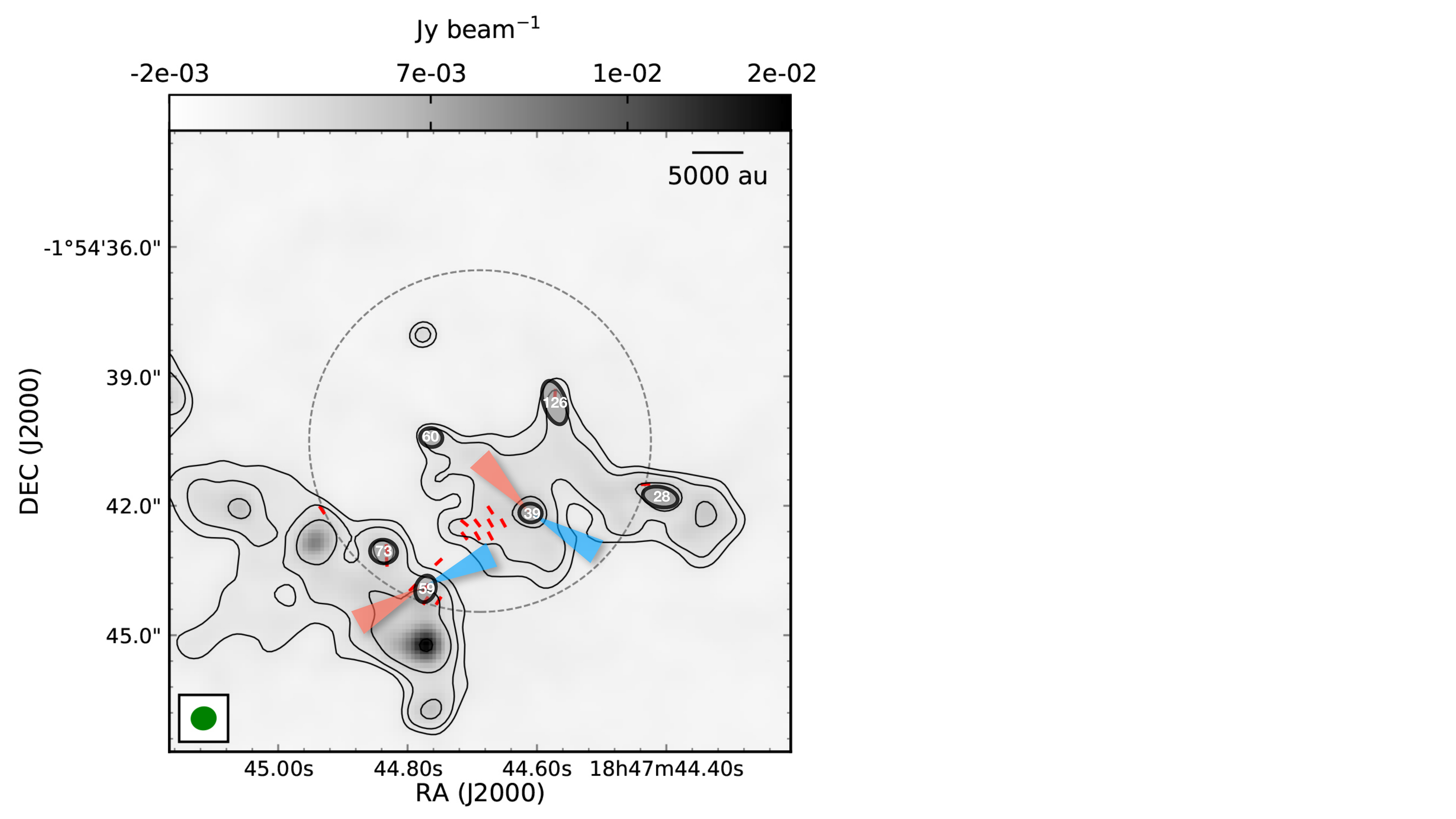}}
    \subfloat[Pointing 5]{\label{fig:5}\includegraphics[trim=2cm 0cm 28.5cm 0cm, width=0.35\linewidth]{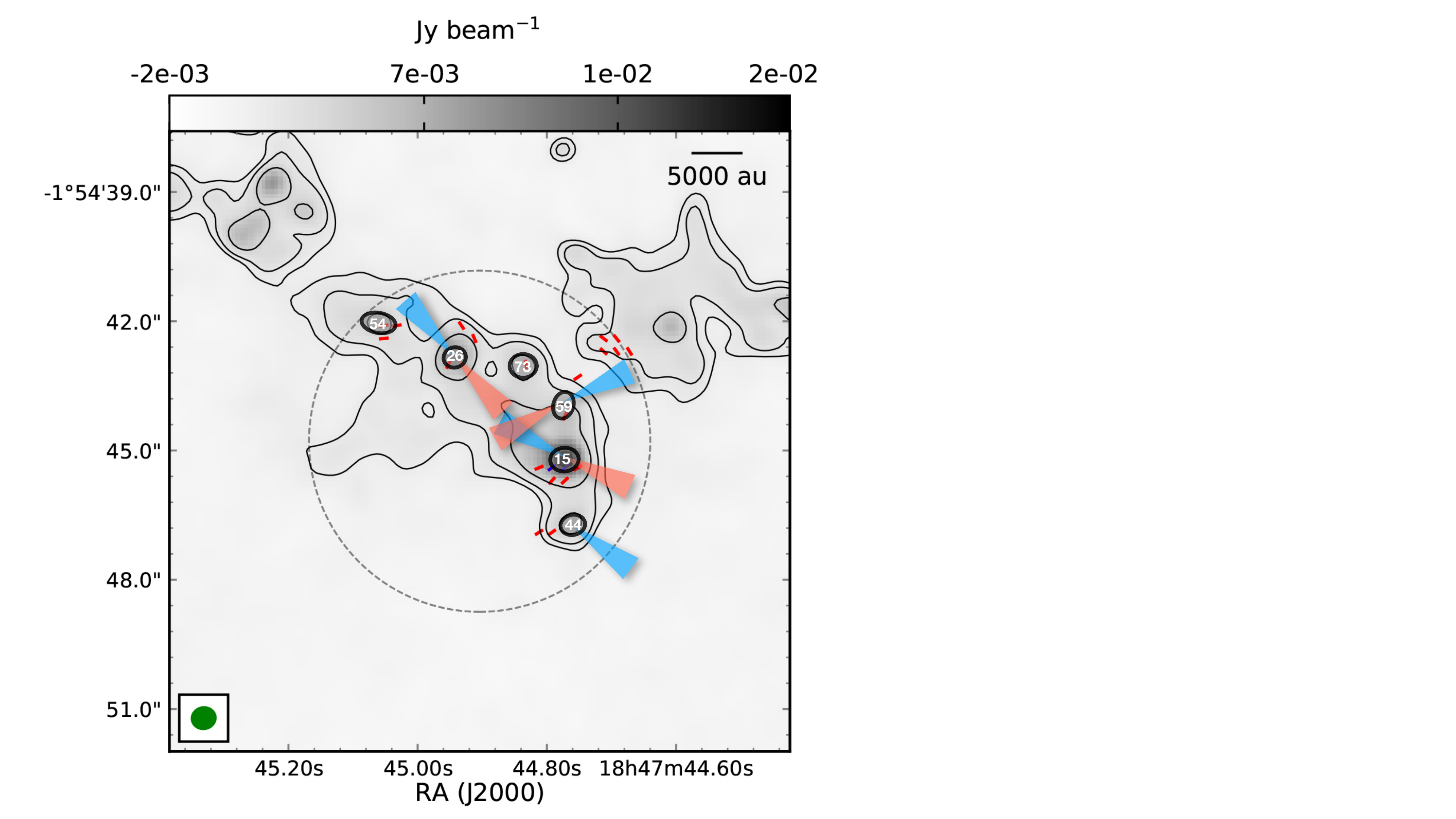}}
\caption{Following Fig.\,\ref{f:pointings_1}: Dust polarization semi-vectors (rotated by 90$^\circ$ to show the inferred magnetic field orientation) over the continuum emission at 1.3\,mm for pointings 4 and 5. Continuum contours levels are at 3, 5, 10, and 40~$\sigma_{\rm I}$ ($\sigma_{\rm I}$ levels in Table~\ref{t:rms}).}
\label{f:pointings_2}
\end{figure*}

We present five pointings with ALMA in full polarization targeting the most massive dense cores in W43-MM1, which were previously identified by \cite{louvet+14} based on IRAM/Plateau de Bure data. The ALMA observations were obtained between April and May 2016 using an array with 35 antennas. We used the standard frequency setup for continuum polarization in band\,6 (i.e., four spectral windows centered at 224.984, 226.984, 239.015, and 241.015\,GHz). Each spectral window has 64 channels of 31.250 MHz each, corresponding to a total bandwidth of 1.875\,GHz per spectral window. The maximum recoverable scale is $\sim$10.6$^{''}$. The resulting continuum images have an equivalent frequency of 233\,GHz (or 1.3\,mm) with an angular resolution of 0.55$^{''}$\,$\times$\,0.49$^{''}$ (or $\sim$2700\,au, considering the distance of the target). The calibration, the imaging, and the analysis were done using version 5.1.2. of the Common Astronomical Software Applications \citep[{\tiny CASA},][]{casa}. 

\begin{table*}[!ht]
\caption{Central coordinates in J2000 and noise values for each pointing. We also show the noise values of the linear polarization maps ($\sigma_{\mathrm{P}}$), which we used to define the 3$\sigma_{\mathrm{P}}$ and 5$\sigma_{\mathrm{P}}$ thresholds to plot the polarization vectors for the individual pointings. Since we did not analyze the polarized emission in the mosaic, we only present its Stokes\,\textit{I} noise level. The noise values are in units of mJy/beam.}              
\label{table:params}      
\centering                                      
\begin{tabular}{ccccccc} 
\hline\hline                        
Pointing  &  RA & Dec & $\sigma_{\mathrm{I}}$ & $\sigma_{\mathrm{Q}}$ & $\sigma_{\mathrm{U}}$ & $\sigma_{\mathrm{P}}$ \\    
\hline\hline                               
1 & 18:47:47.00 & -1:54:26.90 & 3.42 & 0.12 & 0.11 & 0.12 \\

2 & 18:47:46.47 & -1:54:32.10  & 2.51 & 0.11 & 0.10 & 0.11 \\
3 & 18:47:46.51 & -1:54:21.00 &  2.37 & 0.12 & 0.11 &  0.11\\

4 & 18:47:44.68 & -1:54:40.50  &  0.47 & 0.04 & 0.05 &  0.05 \\


5 & 18:47:44.90 & -1:54:44.80  &  0.50 & 0.05 & 0.05 & 0.05 \\

Mosaic & - & - & 2.39 & - &  - &   \\
\hline\hline
\end{tabular}
\label{t:rms}
\end{table*}

To improve the signal-to-noise ratio (S/N) of our images, we performed a phase self-calibration on each of the five calibrated datasets. The images were improved iteratively through four rounds of phase-only self-calibration using the Stokes\,\textit{I} image as a model. The final imaging was performed with the \textit{tclean} task from {\tiny CASA} using a Hogbom deconvolver and a Briggs weighting scheme with a robust parameter of 0.5. The Stokes\,\textit{U} and \textit{Q} maps were imaged individually for each of the five fields with the \textit{tclean} task using the same parameters as above. After the cleaning, a primary beam correction was applied to each map. Table\,\ref{table:params} shows the coordinates for each pointing and the root mean square (rms) noise level for each Stokes parameter. To provide an overview of the region, we present in Figure\,\ref{f:overview} a mosaicked Stokes\,\textit{I} image\footnote{The mosaic was produced with data that were not self-calibrated.}.

The magnetic field maps were computed using the Stokes\,\textit{Q} and \textit{U} intensity maps for each pointing. For this, we created the linear polarized intensity map, $P$\,=\,$\sqrt{Q^2 + U^2}$. However, even when the Stokes\,\textit{Q} and \textit{U} maps could have negative values, $P$ is always positive. In order to correct for this, we applied the debiasing method proposed by \citet{vaillan06}, which employs a Bayesian approach instead of the classical frequentist method, and gives a better estimation of the polarized emission for low signal-to-noise measurements, $\lesssim$\,3$\sigma_{\mathrm{P}}$, where $\sigma_{\mathrm{P}}$ is the rms noise level in the polarization maps. We wielded the 3$\sigma_{\mathrm{P}}$ debiased intensity value as a threshold to select the pixels used to derive the polarization angle maps, which were computed as $\chi=0.5 \arctan(U/Q)$. Finally, we assumed alignment of the grains with respect to the magnetic field due to the "radiative alignment torques" mechanism (RATs), through which an external radiation field causes the dust grain to spin-up, contributing to an efficient alignment between its angular momentum and the magnetic field line \citep{laz07,hoang_laz09, andersson15}. As a consequence, the aligned grains emit thermal radiation, which is polarized perpendicular to the magnetic field, letting us infer the magnetic field morphology onto the plane of the sky by rotating the polarization angle by 90$^{\circ}$.


\section{Results and analysis}
\label{s:results}

Figure\,\ref{f:overview} shows the 1.3~mm continuum emission of W43-MM1, which hosts two clusters of protostars separated by $\sim$0.9\,pc: the main region (pointings 1, 2, and 3) and the southwest (SW) region (pointings 4 and 5). The continuum emission shown here is very similar to that previously presented by \cite{cortes+16} and \cite{mottenat18}. The latter identified 131 cores in the two clusters. Those falling within the one-third area of the primary beams of our five pointings are superposed over Fig.~\ref{f:overview}. Figures\,\ref{f:pointings_1} and \ref{f:pointings_2} display the morphology of the magnetic field in each of the five pointings, which were imaged independently. In the main region, the large-scale magnetic field ($\sim$1\,pc) shows a smooth and ordered morphology. 
In the SW region, the information about the magnetic field orientation is coarser: There are only eight independent locations where the magnetic field orientation could be derived. Linking the large-scale morphology of the magnetic field to its small-scale features goes beyond the scope of this work, and is the subject of a forthcoming publication (Louvet et al. in prep). 
We measured the position angle (PA) of the magnetic field at the location of the cores identified by \cite{mottenat18}. To ensure reliable linear polarization measurements, and following ALMA recommendations, we restrained our analysis to within one-third of the primary beam for each pointing. In total, 52 cores out of the 130 cores of \cite{mottenat18} fall into these trusted areas. Of these 52 cores, 29 display polarized thermal dust emission. We computed the magnetic field orientation for each core as $BF=0.5 \arctan(U/Q)+\pi/2$ (see Sect.~\ref{s:obs}), where \textit{Q} and \textit{U} are the mean Stokes\,$Q$ and $U$ intensities averaged within each core, whose area is defined by a 2D Gaussian fit of its Stokes\,$I$ emission. The polarization angle uncertainties were estimated through error propagation as $\sigma_{\mathrm{\chi}} = 0.5 \times (\sigma_{P}/P)$, where $P$ is the mean polarized intensity within the corresponding core and $\sigma_{\mathrm{P}}$\,=\,$\sqrt{(<Q> \times \sigma_{\mathrm{Q}})^2 + (<U> \times \sigma_{\mathrm{U}})^2 }/ P $ is the rms of the linearly polarized emission.

We compare the magnetic field orientation at the scale of the cores with archival data that provide the orientation of the outflows driven by the cores \citep{nony20} and the PA of the major axis of the cores \citep{mottenat18}. The absolute values of these three sets of angle orientation, together with the relative orientations among the magnetic field lines at the location of the cores, the outflow orientations, and the PA of the cores are presented in Table~\ref{table:angles}. These differences are represented in the form of CDFs in Fig.\,\ref{f:cdfs} and compared with Monte Carlo simulations to see whether there is a favored orientation. These simulations select 100,000 pairs of random 3D vectors with an angle difference within a given range. These vectors are then projected onto the plane of the sky to measure their apparent angle difference. We considered the following five ranges of 3D angle differences: 0$^{\circ}$\,$-$\,20$^{\circ}$ (parallel); 20$^{\circ}$\,$-$\,50$^{\circ}$; 50$^{\circ}$\,$-$\,70$^{\circ}$; 70$^{\circ}$\,$-$\,90$^{\circ}$ (perpendicular); and 0$^{\circ}$\,$-$\,90$^{\circ}$(random). To further investigate the distributions' tendencies, we performed Kolmogorov-Smirnov\,(KS) tests between the observed and the simulated populations. The KS test is a nonparametric test that quantitatively evaluates the difference between the cumulative distributions of two data sets. We chose the KS test over the chi-squared test as the former is preferable to compare nonequally sampled data. Thus, we computed the p-value ($p$) that evaluates if our data and a given synthetic population are drawn from the same parent distribution, with low values of $p$ corresponding to different populations. Following a conservative rule of thumb, we rejected the hypothesis that the two populations are drawn from the same distribution when $p$\,$<$\,0.05.

\begin{table*}[htbp!]
   \caption[]{The acronym PA$_{\mathrm{core}}$ is the cores (major axis) position angle. We note that $\theta_{blue}$ and $\theta_{red}$ are the blue- and red-shifted outflow position angles, respectively. BF is the magnetic field orientation at the location of the cores. The five last columns show the angle differences between the orientations given in column two, three, four, and five.}
   \label{table:angles}
    \small
    \centering     
    \begin{tabular}{ c | c |  c  c | c | c c c | c c}
    \hline
Core    &  PA$_{\mathrm{core}}^{1,2,3}$ & $\theta_{\mathrm{red}}^{1}$            & $\theta_{\mathrm{blue}}^{1}$     & BF$^{1,4}$            &$|BF-PA_{\mathrm{core}}|^{5}$        &$|BF-\theta_{\mathrm{red}}|^{5}$    & $|BF-\theta_{\mathrm{blue}}|^{5}$        & $|PA_{\mathrm{core}}-\theta_{\mathrm{red}}|^{5}$    & $|PA_{\mathrm{core}}-\theta_{\mathrm{blue}}|^{5}$        \\
\hline                                                                                                                                            
1       & $-$56  $\pm$ 4     &     32 $\pm$ 0.5        &                       &    71 $\pm$ 4       & 53 $\pm$ 4          & 39 $\pm$ 3   &                           & 88 $\pm$ 3    &             \\
2       & $-$75  $\pm$ 8     &     25 $\pm$ 0.9        &$-$165 $\pm$ 0.8       & $-$66 $\pm$ 4       &  9 $\pm$ 6          & 89 $\pm$ 3   & 81 $\pm$ 3                & 80 $\pm$ 6    & 90 $\pm$ 6  \\
3       &    57  $\pm$ 6     &    164 $\pm$ 0.9        &    25 $\pm$ 0.5       & $-$84 $\pm$ 8       & 39 $\pm$ 7          & 68 $\pm$ 6   & 71 $\pm$ 6                & 73 $\pm$ 4    & 32 $\pm$ 4  \\
4       &    41  $\pm$ 6     & $-$140 $\pm$ 1          &                       & $-$83 $\pm$ 2       & 56 $\pm$ 4          & 57 $\pm$ 2   &                           &  1 $\pm$ 4    &             \\
7       & $-$88  $\pm$ 8     &     64 $\pm$ 1          &$-$115 $\pm$ 0.5       &                     &                     &              &                           & 28 $\pm$ 6    & 27 $\pm$ 6  \\
8       & $-$90  $\pm$ 9     &     99 $\pm$ 2          & $-$72 $\pm$ 3.0       & $-$65 $\pm$ 4       & 25 $\pm$ 7          & 16 $\pm$ 3   &  7 $\pm$ 4                &  9 $\pm$ 7    & 18 $\pm$ 7  \\
9       &    80  $\pm$ 11    &     37 $\pm$ 0.5        &$-$146 $\pm$ 1.0       &    70 $\pm$ 7       & 10 $\pm$ 9          & 33 $\pm$ 5   & 36 $\pm$ 5                & 43 $\pm$ 8    & 46 $\pm$ 8  \\
10      & $-$54  $\pm$ 12    &                         &$-$156 $\pm$ 0.5       & $-$40 $\pm$ 3       & 14 $\pm$ 9          &              & 64 $\pm$ 2                &               & 78 $\pm$ 8  \\
11      & $-$89  $\pm$ 13    &     83 $\pm$ 4          &$-$116 $\pm$ 0.5       & $-$55 $\pm$ 9       & 34 $\pm$ 11         & 42 $\pm$ 7   & 61 $\pm$ 6                &  8 $\pm$ 10   & 27 $\pm$ 9  \\
12      &    76  $\pm$ 9     &     68 $\pm$ 0.5        &                       & $-$88 $\pm$ 3       & 16 $\pm$ 7          & 24 $\pm$ 2   &                           &  8 $\pm$ 6    &             \\
13      & $-$87  $\pm$ 31    & $-$142 $\pm$ 3          &    41 $\pm$ 4.0       & $-$82 $\pm$ 2       &  5 $\pm$ 21         & 60 $\pm$ 3   & 57 $\pm$ 3                & 55 $\pm$ 21   & 52 $\pm$ 22 \\
14      &    73  $\pm$ 31    &                         &   104 $\pm$ 0.5       & $-$60 $\pm$ 2       & 47 $\pm$ 21         &              & 16 $\pm$ 1                &               & 31 $\pm$ 21 \\
15      & $-$88  $\pm$ 12    & $-$115 $\pm$ 0.5        &    62 $\pm$ 2.0       & $-$53 $\pm$ 6       & 35 $\pm$ 9          & 62 $\pm$ 4   & 65 $\pm$ 4                & 27 $\pm$ 8    & 30 $\pm$ 9  \\
16      & $-$20  $\pm$ 12    & $-$123 $\pm$ 1          &    20 $\pm$ 0.5       & $-$56 $\pm$ 2       & 36 $\pm$ 9          & 67 $\pm$ 2   & 76 $\pm$ 1                & 77 $\pm$ 8    & 40 $\pm$ 8  \\
17      &    72  $\pm$ 34    &                         &                       &    74 $\pm$ 3       &  2 $\pm$ 23         &              &                           &               &             \\
18      & $-$72  $\pm$ 5     &                         & $-$65 $\pm$ 0.5       &    81 $\pm$ 8       & 27 $\pm$ 7          &              & 34 $\pm$ 6                &               &  7 $\pm$ 4  \\
19      &    85  $\pm$ 17    & $-$158 $\pm$ 1          &    17 $\pm$ 4.0       &                     &                     &              &                           & 63 $\pm$ 12  & 68 $\pm$ 12 \\
22a     & $-$84  $\pm$ 23    &     69 $\pm$ 2          &$-$125 $\pm$ 0.5       &                     &                     &              &                           & 27 $\pm$ 16  & 41 $\pm$ 16 \\
22b     & $-$84  $\pm$ 23    &    142 $\pm$ 0.5        &                       &                     &                     &              &                           & 46 $\pm$ 16  &             \\
23      & $-$77  $\pm$ 25    &                         & $-$68 $\pm$ 1.0       &    87 $\pm$ 7       & 16 $\pm$ 18         &              & 25 $\pm$ 5                &               &  9 $\pm$ 17 \\
26      & $-$67  $\pm$ 16    & $-$139 $\pm$ 0.5        &    41 $\pm$ 0.5       & $-$23 $\pm$ 8       & 44 $\pm$ 13         & 64 $\pm$ 6   & 64 $\pm$ 6                & 72 $\pm$ 11   & 72 $\pm$ 11 \\
27      &    72  $\pm$ 41    &                         &                       &    37 $\pm$ 2       & 35 $\pm$ 28         &              &                           &               &             \\
28      &    65  $\pm$ 11    &                         &                       &    89 $\pm$ 8       & 24 $\pm$ 10         &              &                           &               &             \\
29      & $-$58  $\pm$ 17    &                         & $-$12 $\pm$ 1.0       &                     &                     &              &                           &               & 46 $\pm$ 12 \\
31      &    40  $\pm$ 22    &    101 $\pm$ 1          & $-$73 $\pm$ 1.0       &                     &                     &              &                           & 61 $\pm$ 15   & 67 $\pm$ 15 \\
33      &    33  $\pm$ 24    &                         &                       & $-$70 $\pm$ 9       & 77 $\pm$ 18         &              &                           &               &             \\
36      &    26  $\pm$ 16    &  $-$75 $\pm$ 0.6        &   105 $\pm$ 2.0       &                     &                     &              &                           & 79 $\pm$ 11   & 79 $\pm$ 11 \\
39      &    86  $\pm$ 27    &     43 $\pm$ 4          &$-$120 $\pm$ 0.9       &                     &                     &              &                           & 43 $\pm$ 19   & 26 $\pm$ 19 \\
40      & $-$79  $\pm$ 23    &                         &                       &    84 $\pm$ 7       & 17 $\pm$ 17         &              &                           &               &             \\
41      & $-$40  $\pm$ 32    &                         &                       &    12 $\pm$ 7       & 52 $\pm$ 23         &              &                           &               &             \\
44      & $-$66  $\pm$ 19    &                         &$-$126 $\pm$ 0.5       & $-$63 $\pm$ 9       &  3 $\pm$ 15         &              & 63 $\pm$ 6                &               & 60 $\pm$ 13 \\
47      & $-$96  $\pm$ 30    &                         &                       & $-$59 $\pm$ 3       & 37 $\pm$ 21         &              &                           &               &             \\
48      &    46  $\pm$ 26    &                         &                       &                     &                     &              &                           &               &             \\
49      &    86  $\pm$ 17    &  $-$45 $\pm$ 0.5        &   150 $\pm$ 0.5       &                     &                     &              &                           & 49 $\pm$ 12   & 64 $\pm$ 12 \\
54      &    81  $\pm$ 12    &                         &                       & $-$79 $\pm$ 7       & 20 $\pm$ 10         &              &                           &               &             \\
59      & $-$14  $\pm$ 33    &    119 $\pm$ 0.5        & $-$65 $\pm$ 0.5       & $-$29 $\pm$ 9       & 15 $\pm$ 24         & 32 $\pm$ 6   & 36 $\pm$ 6                & 47 $\pm$ 23   & 51 $\pm$ 23 \\
60      &    81  $\pm$ 28    &                         &                       &                     &                     &              &                           &               &             \\
67      &    40  $\pm$ 16    &     12 $\pm$ 1.5        &$-$174 $\pm$ 0.5       &                     &                     &              &                           & 28 $\pm$ 11   & 34 $\pm$ 11 \\
73      &    81  $\pm$ 25    &                         &                       &    11 $\pm$ 8       & 70 $\pm$ 18         &              &                           &               &             \\
80      &    70  $\pm$ 23    &                         &                       &                     &                     &              &                           &               &             \\
96      & $-$62  $\pm$ 21    &                         &                       &    14 $\pm$ 5       & 76 $\pm$ 15         &              &                           &               &             \\
117     & $-$83  $\pm$ 12    &                         &                       &                     &                     &              &                           &               &             \\
126     & $-$81  $\pm$ 14    &                         &                       &     3 $\pm$ 8       & 84 $\pm$ 11         &              &                           &               &             \\
 \hline
  \end{tabular} 
    \tablefoot{
$^{1}$: All of the angles were measured counterclockwise from north. \\
$^{2}$: The PAs (major axis) come from the {\tiny GETSOURCES} catalog and complete Table\,1 of \cite{mottenat18}.\\
$^{3}$: To account for the ellipticity of cores, the error of the PA was computed as $180^\circ\times(1-e)\times(\ln[{F_{\rm core}/\sigma_f}])^{-2}$ where $e$ is the ellipticity of the core, $\sigma_f$ is the uncertainty on the flux measurement, and $F_{\rm core}$ is the integrated flux of the core (S. Mensh'chikov, private communication).\\
$^{4}$: The magnetic field PA indicates the mean magnetic field orientation inside the cores (see Sect.~\ref{s:results} for details).\\
$^{5}$: The uncertainty for the angle difference (col. 6, 7, 8, 9, and 10) is $\sigma$\,=\,$\arctan\left(\sqrt{\dfrac{\sin{\delta}\,^2 + \sin{\phi}\,^2}{\cos{\delta}\,^2 + \cos{\phi}\,^2}}\right)$, where $\delta$ and $\phi$ correspond to the errors associated with the measurement uncertainties (col. 2, 3, 4, and 5).
}
  \end{table*}

\subsection{Comparison of the magnetic field orientation with that of the core major axis}
\label{ss:bfvscores}

We compared the PA of the 29 cores falling within the one-third region of the primary beams of our five pointings with the mean PA of the magnetic field inside the cores. The CDF of these differences in angle is presented in Fig.~\ref{f:cdf-corespa_bfield}, together with the CDFs arising from the five different synthetic angle simulations. The CDF of the data clearly shows that the orientation of the cores with respect to the magnetic field orientation is only reproduced by the 20-50$^\circ$ population. The results of the KS tests, which compare the observations with the synthetic populations, are reported in Table~\ref{table:ksres}. Here, we confirm that the magnetic field orientation with respect to the cores PA in our data is not random: All of the synthetic populations (including the random distribution), except for the 20-50$^\circ$ distribution, are rejected by the KS test.

\subsection{Comparison of the magnetic-field orientation with that of the outflows axis}
\label{ss:bfvsflows}

\begin{figure*}[!ht]
    \centering
    \subfloat[ ]{\label{f:cdf-corespa_bfield}\includegraphics[trim=0cm 0cm 0cm 0cm, width=0.33\linewidth]{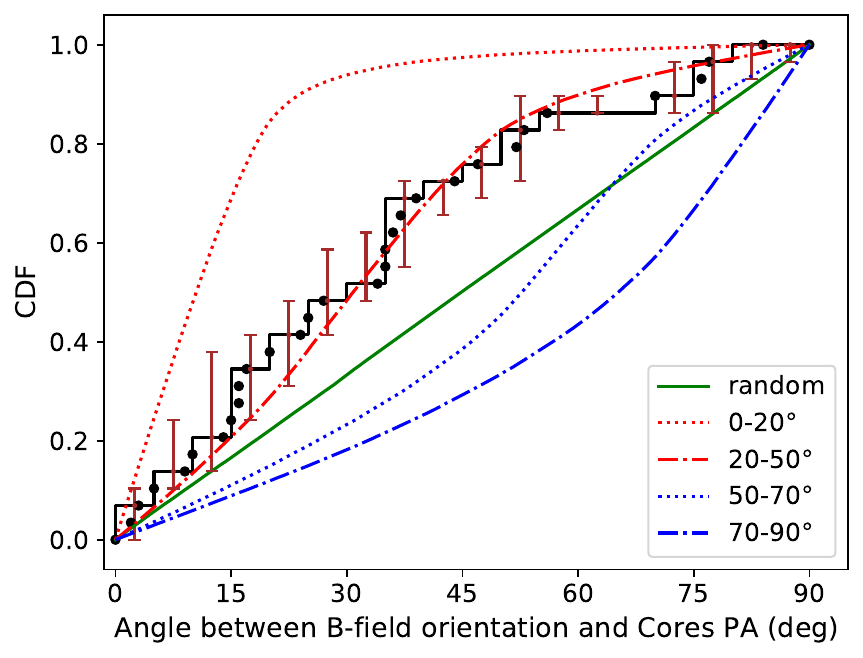}}
    \subfloat[]{\label{f:cdf-bfout}\includegraphics[trim=0cm 0cm 0cm 0cm, width=0.33\linewidth]{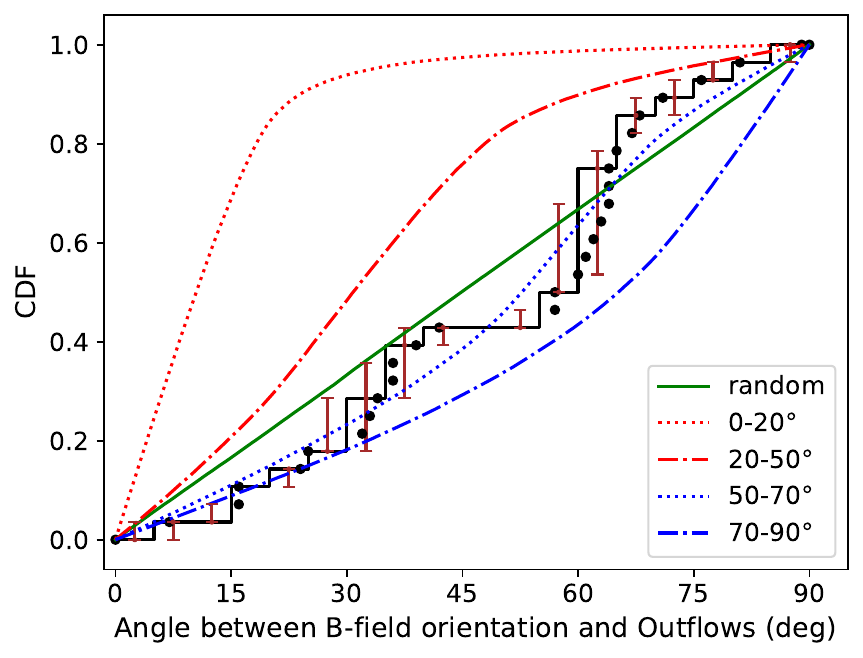}}
    \subfloat[]{\label{f:cdf-corespa_outf}\includegraphics[trim=0cm 0cm 0cm 0cm, width=0.33\linewidth]{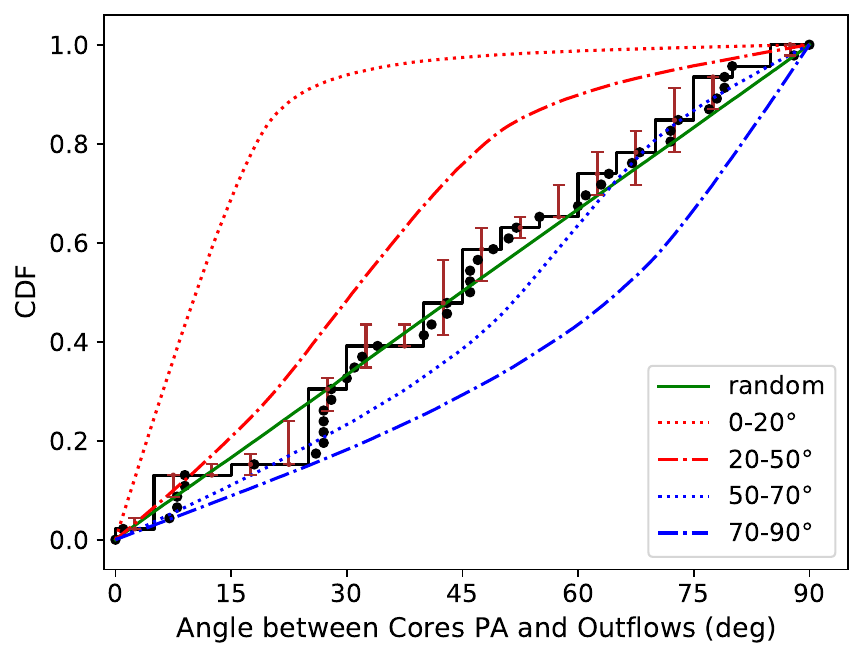}}\\
    \caption{Cumulative distribution function (CDF, black points) of the angle difference between the magnetic field orientation and the PA of the cores (\ref{f:cdf-corespa_bfield}), the magnetic field and the outflows (\ref{f:cdf-bfout}), and the outflows and the PA of the cores (\ref{f:cdf-corespa_outf}) in W43-MM1. The black histogram shows the CDF with 5$^\circ$ bin-widths. The brown bars show the errors on the CDF for each bin. Colored curves are the CDFs resulting from Monte Carlo simulations of different 3D angles projected onto the plane of the sky. The different intervals of angles considered for the synthetic populations are indicated on the bottom right of the plots.}
    \label{f:cdfs}
\end{figure*}

\cite{nony20} studied the outflows associated with the cores of W43-MM1 by observing the $^{12}$CO(2-1) and the SiO(5-4) molecular emission lines with an ALMA plus ACA mosaic at an angular resolution of $\sim$0.45$^{\prime\prime}$, which is similar to the present study. In total, they report 46 individual outflow lobes that are associated with 27 cores. On the one hand, about one-fourth of these outflows are monopolar. They could be truly monopolar, as observed in some nearby objects (e.g., HH30, \citealt{louvet+18}), or the emission arising from the companion outflow could be absorbed by the surrounding medium. On the other hand, most of the bipolar outflows are not straight, meaning that there is a shift between the PA of the red-shifted lobe with respect to the PA of the blue-shifted lobe (see Figs.\,\ref{f:pointings_1} and \ref{f:pointings_2}). The authors suggest that this could be due to subfragmentation within the cores and/or deflection of the outflows. For this reason, we consider each outflow lobe independently\footnote{We also performed the analysis treating bipolar outflows as single data points and obtained similar results as the ones presented in Sect.\,\ref{ss:bfvsflows} and Sect.\,\ref{ss:outfvscores}.}. Eighteen of the 27 cores that have one or more outflows are detected in polarization. In Table~\ref{table:angles}, we report the orientation of the outflows associated with these 18 cores. The orientation of one outflow is defined as the angle of the line linking the core location, which is defined as the continuum peak, and the furthest knot of the outflow (see \citealt{nony20} for more details about the outflow characteristics). 

The CDF of the orientation differences between the magnetic field and the outflows is presented in Fig.~\ref{f:cdf-bfout}. In Table\,\ref{table:ksres} we show the results of the KS tests, which compare the observations with the synthetic populations. Based on these tests, we can reject the hypothesis that the outflows are oriented parallel or perpendicular to the magnetic field. Instead, the KS tests favor either a random orientation of the outflows with respect to the magnetic field lines or the outflows are oriented 50-70$^{\circ}$ with respect to the magnetic field lines.

\subsection{Comparison of the orientation of the outflows with that of the core major axis}
\label{ss:outfvscores}
The comparison between the PA of the outflows and the PA of the cores is presented in the CDF in Fig.\,\ref{f:cdf-corespa_outf}. Based on visual inspection, the observational CDF seems to be consistent with the random synthetic population, suggesting that there is no specific orientation of the outflows with respect to the elongation of the cores. However, when comparing the observations with the synthetic populations through the KS tests, while the random population obtains the highest statistical weight with \textit{p} $\sim$0.5, the 50-70$^\circ$ population cannot be discarded (\textit{p}$\sim$0.2). 

\begin{table}[htbp!]
   \caption[]{Kolmogorov-Smirnov tests for the observed and simulated angle distributions.}
   \label{table:ksres}
    \small
    \centering     
    \begin{tabular}{c|c|c|c} 
    \hline
Test case                 & Synthetic population                         & KS $p$-value        & Sample size         \\ 
\hline 
                          & 0-20$^\circ$                                      & 7\,$\times$\,10$^{-8}$                                     \\
                          & 20-50$^\circ$                                     & 0.76                                     \\
B-field vs. Cores PA     & 50-70$^\circ$                                     & 6\,$\times$\,10$^{-3}$         & 29                             \\
                          & 70-90$^\circ$                                     & 6\,$\times$\,10$^{-6}$                                 \\
                          & Random                                            & 4\,$\times$\,10$^{-2}$                                    \\

 \hline
                          & 0-20$^\circ$                                      & 2\,$\times$\,10$^{-100}$          \\
                          & 20-50$^\circ$                                     & 8\,$\times$\,10$^{-6}$                                      \\
B-field vs. Outflows      & 50-70$^\circ$                                     & 0.51               & 28                  \\
                          & 70-90$^\circ$                                     & 1\,$\times$\,10$^{-2}$                                     \\
                          & Random                                            & 0.16                                    \\
\hline  

                          & 0-20$^\circ$                                      & 1\,$\times$\,10$^{-15}$                                      \\
                          & 20-50$^\circ$                                     & 1\,$\times$\,10$^{-3}$                                      \\
Cores PA vs. Outflows     & 50-70$^\circ$                                     & 0.16               & 46               \\
                          & 70-90$^\circ$                                     & 1\,$\times$\,10$^{-3}$                                      \\
                          & Random                                            & 0.53                                     \\
\hline                                                                                                                        

  \end{tabular} 
  \end{table}


\section{Discussion}
\label{s:diss}
Our observations recover an ordered magnetic field pattern in the main protocluster of the W43-MM1 HMSFR (see Fig.\,\ref{f:pointings_1}), plus a few detections in its SW protocluster (see Fig.\,\ref{f:pointings_2}). This is in perfect agreement with previous observations of polarized dust emission toward the W43-MM1 HMSFR \citep{cortes-crutcher,sridha+14,cortes+16}. In the following subsections, we discuss the direction of the magnetic field compared to the orientation of the cores and outflows.

\subsection{The alignment of the cores with respect to the magnetic field}
\label{ss:corevsbf}
Magnetic fields may contribute to support clouds against gravity, thus indirectly affecting the evolution of individual cores. Furthermore, magnetic fields could also directly affect the evolution of cores. In the so-called magnetically regulated core-collapse models \citep[e.g.,][]{galli93a,galli93b, allen+03b,allen+03a}, magnetic fields are dynamically important and dominate the dynamics of the core-collapse by deflecting the infalling gas toward the equatorial plane to form a flattened structure, known as a pseudodisk, around the central protostar. The pseudodisks are not supported by rotation but by magnetic pressure and they develop orthogonally to the magnetic field direction. They have sizes of a few thousands of astronomical units, on the order of the angular scales probed in the present study. Therefore, if magnetically regulated core-collapse were at work, one would expect to observe flattened central regions -- the pseudodisks -- which are generically referred to as cores in the present study and oriented perpendicular to the magnetic field lines. The very few existing observational studies, which have reported this type of behavior, had very limited samples: \citet{davidson11} studied 350\,$\mu$m polarization observations taken at the CSO toward three low-mass Class 0 protostars, and they report a magnetic field orientation, which is in loose agreement with the magnetically-controlled core-collapse predictions; \citet{chapman13} found the magnetic field orientation to be perpendicular to the pseudodisks in four low-mass cores, using 350\,$\mu$m polarization observations taken at the CSO; \cite{qiu+14} report a similar result toward the G240.31+0.07 high-mass core upon using the SMA at 0.88\,mm. In contrast, our results, which were obtained with a sample of 29 cores, are not consistent with the prediction of the magnetically-regulated core-collapse models. Instead, as is shown in Fig.\,\ref{f:cdf-corespa_bfield}, the major axes of the cores are not preferentially oriented orthogonal to the magnetic field, but rather they tend to be oriented 20-50$^\circ$ with respect to the field. Having a nonrandom orientation of the cores with respect to the magnetic field orientation demonstrates that the magnetic field and the dense material are well coupled.

It is interesting to point out that ideal MHD simulations of gravitational collapse of cores have shown that such configurations, where the rotation axis of the core is not parallel to the magnetic field direction, reduce the magnetic braking and favor the formation of circumstellar disks \citep[e.g.,][]{joos12}. More specifically, recent nonideal MHD simulations have reported that such configurations would engender warped disks, implying that the PA of the pseudo-disks would differ from the PA of the inner disks \citep[see, e.g., the Fig.\,3-middle row of][]{tsukamoto+18}.

\subsection{The orientation of the outflows with respect to the magnetic field}
\label{ss:flotvsbf}

One common feature of the two families of models explaining the presence of outflows -- the disk-wind models \cite[see, e.g.,][]{ferreira97} and the entrainment models \cite[see e.g.,][]{blandford82} -- is that they both need the circumstellar disk to be orthogonal to the magnetic field lines, causing the ejection of matter along the magnetic field lines. Hence, if the models are valid, one expects to see alignment between the outflows and the magnetic field near the disks. \cite{carrasco-gonzalez+10} first found an alignment between the outflow-lobe associated to the massive YSO HH 80/81 and the magnetic field in its jet, inferred from polarized synchrotron emission measured with the VLA, up to $\sim$0.5\,pc from the driving source. \cite{beuther10} studied the high-mass disk-outflow system IRAS18089-1732 and found that the magnetic field, which was measured from CO(3-2) polarized emission, is aligned with the outflow orientation from small core scales ($\sim$7000\,au) to larger outflow scales of $\sim$36000\,au. \cite{sridha+14} also reported alignment between the magnetic field, which was measured from dust polarized emission, and one outflow in W43-MM1 by using SMA observations at an angular resolution of $\sim$3$^{''}$. However, this interpretation has since been disproved by a more recent study using ALMA at $\sim$0.5\,$^{''}$, which revealed that the outflow seen with the SMA toward W34-MM1 was the sum of 12 outflow lobes with different orientations \citep{nony20}. Also, \cite{zhang14} studied a sample of 14 clumps located in different regions where they could compare the orientation of the magnetic field lines with respect to the axes of the outflows. While they refrained from interpreting the detailed structure of the obtained distribution, they report no strong correlation between the outflow and the magnetic field orientations. The lack of alignment between outflows and the magnetic field is also observed in low-mass star-forming regions at the $\sim$1000\,au scale \citep[see the recent review by][]{hullrev_19}. At the envelope scale ($\sim$600-1500\,au), \cite{galametz18} report that the magnetic field is either preferentially observed aligned or perpendicular to the outflow direction when studying a sample of 12 Class\,0 envelopes in nearby clouds. They interpret this bimodality by considering that the cases that showed an alignment between the magnetic field and the rotation axis might be the result of a strong magnetic field. They also attribute this bimodality to the fact that most of the cores with a magnetic field perpendicular to the outflows are binaries.

In this work, which is based on a sample of 28 outflow lobes, we find two configurations to be compatible with the observational CDF of the angle difference between the orientation of the outflows and the magnetic field orientation (see Sect.~\ref{ss:flotvsbf}).

The outflows could be randomly oriented to the magnetic field lines, which is in line with the observations toward low-mass star-forming regions \citep{hullrev_19}. Assuming that the models for outflow launching are accurate, where gas from the accretion disk should be ejected along magnetic field lines, a random orientation would imply that most of our (high-mass) protostars form in an environment where the magnetic field is too weak to maintain a consistent orientation from the $\sim$2700\,au scales that we are probing, down to the 0.1-10\,au scales where outflows are launched \citep{louvet+18}. It would also imply that the orientation of the disks is not controlled by the magnetic field, but by another mechanism provoking angular momentum redistribution, such as interactions in multiple systems or the randomization of the disk orientation by accretion during the early phases of protostellar evolution \citep{lee17}.

Alternatively, the outflows could be oriented 50-70$^\circ$ to the magnetic field lines. It is striking to note that if the elongation of the cores corresponds to the major axes of the underlying disks, and considering that outflows propagate orthogonally to their hosting disk, a population of cores that is oriented 20-50$^\circ$ as reported in Sect.~\ref{ss:corevsbf} would translate into a population of outflows oriented 40-70$^\circ$ with respect to the magnetic field lines, which is in excellent agreement with the outflow being preferentially oriented 50-70$^\circ$ to the magnetic field. It is this coherent view, arising from independent correlations, that we might be witnessing here. Nevertheless, we acknowledge that there is a better match for the CDF presenting the 20-50$^\circ$ angle differences between the PA of the cores and the magnetic field orientations than the CFD presenting the 50-70$^\circ$ angle differences between the outflow orientations and the magnetic field. Such a loss of statistical significance is expected if, in some cores, the magnetic field is not strong enough to govern the angular momentum from the core scale down to the outflow ejection location at 0.1-10\,au.


\subsection{The orientation of the outflows with respect to their hosting cores}
\label{ss:corevsflot}

The analysis presented in Sect.~\ref{ss:outfvscores} asserts that two configurations are consistent with our data, that is, either the outflows are randomly oriented or oriented 50-70$^\circ$ to their hosting cores. We discuss in turn these two possibilities.

We first comment on the possibility of a  random orientation of the outflows with respect to their hosting cores. Recent studies of protostellar objects forming low-mass stars have shown that outflows are usually observed perpendicular to their underlying disks at small scales \citep[$<$200\,au, see e.g.,][]{louvet+16disk}. Building on the hypothesis that this also happens for high-mass protostars, there are two ways of interpreting this result. If the core elongations are representative of the major axes of underlying disks, the random orientation of the outflows to the cores would indicate that the outflows get randomly deflected at scales below our angular resolution ($\sim$2700\,au). Such deflections have been observed in a few outflows in W43-MM1 at a larger scale by \cite{nony20} and, therefore, could also occur below our resolution. Nevertheless, it is reasonable to argue that a deflection is more likely to occur with a small angle and that the statistical impact of deflections should be minimal. Therefore, deflections are unlikely to result in a random orientation of the outflows with respect to their hosting cores. If outflows are indeed randomly oriented with respect to the core PA, a more likely explanation is that the elongation of the cores is not representative of the PA of underlying disks. In such a case, the angular momentum of the disks is set by processes acting at scales smaller than the resolution of our observations.

Commenting on the  possibility that outflows are oriented 50-70$^\circ$ to their hosting cores, it is important to note that this type of configuration does not match any expectations from the models of launching and/or propagation of outflows. In order to come up with a more physical explanation, we tested bimodal synthetic populations. The fiducial synthetic distributions of angles presented in Sect.~\ref{s:results} only contain single populations, meaning, for example,  that in our "50$^\circ$-70$^\circ$" distribution all of the angle differences are between 50$^\circ$ and 70$^\circ$ in the 3D space. Instead, the bimodal synthetic populations have a percentage $n_{\%}$ of 3D vector pairs with angle differences within a certain angle range and the complementary fraction of vector pairs, $1-n_{\%}$, within another angle range. We found a good correlation of the observations with a bimodal population where 85\% of the outflows are randomly oriented and 15\% are orthogonally oriented to their respective core, which is in line with the assumption from Sect.~\ref{ss:flotvsbf}. However, given the size of our sample, it is not statistically possible to discriminate between the 50-70$^\circ$ orientation, a random orientation, or an 85\% random population plus a 15\% orthogonal population.


\section{Conclusions}\label{conc}

We report thermal dust polarized emission toward a sample of dense cores in the high-mass protostellar cluster W43-MM1, at an angular resolution of $\sim$0.50$^{\prime\prime}$ ($\sim$\,0.01\,pc, or 2700\,au), using ALMA observations in Band\,6. We compare the orientation of the magnetic field with archival data of the orientation of 29 dense cores and 28 outflow lobes in this region.

The major axes of the cores are not randomly oriented with respect to the magnetic field, showing that the magnetic field is well coupled to the dense material composing the cores. Instead, the cores are compatible with an orientation of 20-50$^\circ$ with respect to the magnetic field. If confirmed, this result rules out the magnetically-controlled core-collapse models in which a flattened envelope, or pseudodisk, is expected to develop orthogonally to the magnetic field lines. The outflows are oriented 50-70$^\circ$ with respect to the magnetic field (i.e., the orientation of the cores plus 90$^\circ$) or randomly oriented. Given our statistics, we could not discriminate between these two possibilities.

We propose that, in some cases, the magnetic field at the scale of cores is strong enough to set the orientation of the disk, resulting in an outflow versus magnetic field orientation which is coherent with that of the core versus magnetic field. In other cases, the magnetic field at the scale of cores is not strong enough to set the orientation of the disk. The latter is then controlled by other mechanisms, such as angular momentum redistribution and/or gravitational interaction in multiple systems, which results in a random orientation of the outflow with respect to the core scale magnetic field orientation.

\begin{acknowledgements}
      \noindent{We thank the anonymous referee for her/his helpful comments that helped us clarify the article. FKL thanks Maud Galametz and Sylvie Cabrit for fruitful discussions about the collapse of cores and the propagation of outflows, respectively. FKL acknowledges the support of the Fondecyt program Nº\,3170360. FM acknowledges the support of the Joint ALMA Observatory Visitor Program. C.L.H.H. acknowledges the support of both the NAOJ Fellowship as well as JSPS KAKENHI grants 18K13586 and 20K14527. This paper makes use of the following ALMA data: ADS/JAO.ALMA\#2015.1.01020.S. ALMA is a partnership of ESO (representing its member states), NSF (USA) and NINS (Japan), together with NRC (Canada), MOST and ASIAA (Taiwan), and KASI (Republic of Korea), in cooperation with the Republic of Chile. The Joint ALMA Observatory is operated by ESO, AUI/NRAO and NAOJ. This work was supported by the Programme National de Physique Stellaire and Physique et Chimie du Milieu Interstellaire (PNPS and PCMI) of CNRS/INSU (with INC/INP/IN2P3) co-funded by CEA and CNES. GG and LB acknowledge support from CONICYT project AFB 170002.}
\end{acknowledgements}

\bibliographystyle{aa} 

\bibliography{main.bib}

\end{document}